\documentclass[useAMS,usenatbib]{mn2e}
\usepackage{graphicx}
\title[Parsec-scale SiO Emission in an IRDC]{Parsec-scale SiO Emission in an Infrared Dark Cloud}

\author[I. Jim\'enez-Serra et al.]{I. Jim\'{e}nez-Serra$^{1,2}$\thanks{E-mail: i.jimenez-serra@cfa.harvard.edu}, P. Caselli$^{1}$, J. C. Tan$^{3}$, A. K. Hernandez$^{3}$, F. Fontani$^{4}$,  
\newauthor M. J. Butler$^{3}$ and S. van Loo$^{1}$ \\
$^{1}$School of Physics \& Astronomy, E.C. Stoner Building,
University of Leeds, Leeds, LS2 9JT, UK\\
$^{2}$Harvard-Smithsonian Center for Astrophysics, 60 Garden St., 02138, Cambridge, MA, USA \\
$^{3}$Department of Astronomy, University of Florida, Gainesville, FL 32611, USA\\
$^{4}$Institut de Radioastronomie Millim\'etrique, 300 rue de la Piscine, 38406 St. Martin d'Heres, France}

\begin{document}

\date{Accepted 1988 December 15. Received 1988 December 14; in original form 1988 October 11}

\pagerange{\pageref{firstpage}--\pageref{lastpage}} \pubyear{2002}

\maketitle

\label{firstpage}

\begin{abstract}

We present high-sensitivity 2$'$$\times$4$'$ maps of the 
$J$=2$\rightarrow$1 rotational lines of SiO, CO, $^{13}$CO and C$^{18}$O, 
observed toward the filamentary Infrared Dark Cloud (IRDC) G035.39-00.33. Single-pointing
spectra of the SiO $J$=2$\rightarrow$1 and $J$=3$\rightarrow$2 
lines toward several regions 
in the filament, are also reported. The SiO images reveal  
that SiO is widespread along the IRDC (size $\geq$2$\,$pc), showing two different 
components: one bright and compact arising from three condensations (N, E and S),
and the other weak and extended along the filament. While the first component shows
broad lines (linewidths of $\sim$4-7$\,$km$\,$s$^{-1}$) in both SiO 
$J$=2$\rightarrow$1 and SiO $J$=3$\rightarrow$2, 
the second one is only detected in SiO $J$=2$\rightarrow$1 and has narrow lines 
($\sim$0.8$\,$km$\,$s$^{-1}$). The maps of CO and its isotopologues 
show that low-density filaments are intersecting the IRDC and appear 
to merge toward the densest portion of the cloud. This resembles
the molecular structures predicted by flow-driven, shock-induced and 
magnetically-regulated cloud formation models. As in outflows associated
with low-mass star formation, the excitation temperatures and fractional
abundances of SiO toward N, E and S, increase with velocity 
from $\sim$6 to 40$\,$K, and from $\sim$10$^{-10}$ to $\geq$10$^{-8}$ 
respectively, over a velocity range of $\sim$7$\,$km$\,$s$^{-1}$. 
Since 8$\,$$\mu$m sources, 24$\,$$\mu$m sources and/or extended 
4.5$\,$$\mu$m emission are detected in N, E and S, 
broad SiO is likely produced in outflows associated with high-mass 
protostars. The excitation 
temperatures and fractional abundances of the narrow SiO lines, however,  
are very low ($\sim$9$\,$K and $\sim$10$^{-11}$,
respectively), and consistent with the processing of interstellar grains
by the passage of a 
shock with $v_s$$\sim$12$\,$km$\,$s$^{-1}$. 
This emission could be generated i) by a large-scale shock, perhaps remnant of the IRDC 
formation process; ii) by decelerated or recently processed gas in large-scale 
outflows driven by 8$\,$$\mu$m and 24$\,$$\mu$m sources; 
or iii) by an undetected and widespread population of lower mass 
protostars. High-angular resolution observations are needed to disentangle between 
these three scenarios.  

\end{abstract}


\section{Introduction}

Infrared Dark Clouds (IRDCs) are high-extinction regions viewed against the
diffuse mid-IR Galactic background \citep[][]{per96,egan98}. These clouds are cold
\citep[$T<$25$\,$K;][]{pill07} and exhibit a range of densities from  
$n$(H)$\geq$10$^3$$\,$cm$^{-3}$ to $\geq$10$^4$-10$^5$$\,$cm$^{-3}$
in their clumps and cores \citep[][]{tey02,but09}. Since these structures have masses and mass surface densities 
similar to regions that are known to be forming massive
protostars and star clusters, they may represent the initial conditions for
massive star and star cluster formation \citep[Rathborne, Jackson \& 
Simon 2006;][]{zha09,rag09}.

It is well-known that silicon monoxide (SiO) is an excellent tracer 
of molecular gas processed by shocks. While SiO
is heavily depleted onto dust grains in the 
quiescent gas of dark clouds such as L183 \citep[upper limits of the SiO 
fractional abundance of $\leq$10$^{-12}$; Ziurys, Friberg \& Irvine 1989;][]{req07}, this molecule is enhanced by large 
factors (in some cases by $>$10$^6$) toward molecular outflows \citep[][]{mar92}.
This is due to the injection of molecular material into the gas phase 
by the processing of dust grains \citep[e.g.][]{cas97,sch97,jim08,gui07,gui09}. 

The typical SiO emission measured in molecular outflows  
shows broad line profiles with linewidths of some tens of 
km$\,$s$^{-1}$ \citep[][]{mar92}. Narrower SiO lines
have also been detected toward low-mass star forming regions such 
as NGC1333 and L1448-mm \citep[][]{lef98,jim04}. Although the
nature of this emission is not clear yet, \citet{lef98} have 
proposed that these lines could trace shocked material 
deflected and decelerated by the interaction with 
pre-existing clumps. Alternatively, narrow SiO could arise from gas 
recently processed by the magnetic precursor of 
young magneto-hydrodynamic (MHD) shocks \citep{jim04}. 
 
In the case of IRDCs, \citet{mott07} and \citet{beu07} have recently carried out 
two large surveys of SiO emission toward the Cygnus X molecular cloud complex and 
toward a sample of IRDCs, respectively. Their single-pointing observations show that 
the detection rate of SiO toward IR-quiet massive cores close to sites
of on-going star formation, is relatively high. This is expected since
SiO is tightly associated and restricted to shocked gas in outflows.  
However, no widespread narrow SiO emission, as seen toward NGC1333, has been reported 
in IRDCs so far.

We present the detection of widespread SiO $J$=2$\rightarrow$1 
emission (size of $\geq$2$\,$pc) toward the IRDC G035.39-00.33
\citep[Cloud H in][]{but09}. From its filamentary morphology, this IRDC 
is believed to be at the early stages of its evolution as predicted by dynamical 
models of giant molecular cloud formation \citep[][Heitsch, Stone \&
Hartmann 2009]{van07,hen08}. 
The observed large-scale SiO feature is probably a composition of 
broad and compact emission, linked to outflows associated with
high-mass star formation; and extended narrow SiO emission. 
The observed narrow SiO lines could be explained i) by a large-scale shock, remnant of the IRDC formation process; ii) by 
decelerated or recently processed gas in the precursor of MHD shocks in 
large-scale outflows, probably driven by the 8$\,$$\mu$m and 24$\,$$\mu$m sources
observed in the IRDC; or iii) by an undetected and widespread population of 
lower-mass protostars. 

\section{Observations and Results}
\label{res}

\begin{figure*}
\begin{center}
\includegraphics[angle=270,width=1.0\textwidth]{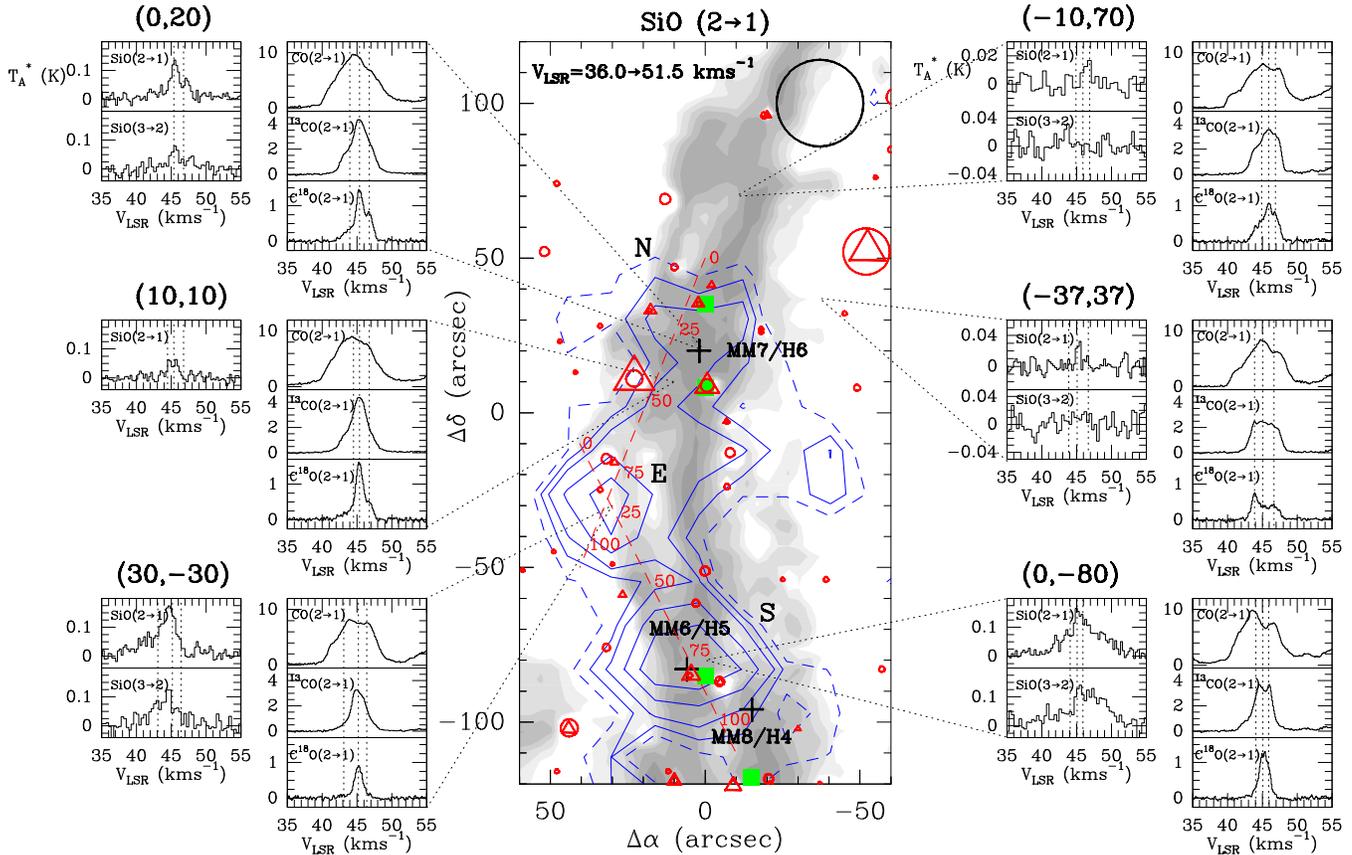}
\caption{{\it Central panel}: Integrated intensity map of the SiO $J$=2$\rightarrow$1 line 
toward G035.39-00.33 for the velocity range from 36 to 
51.5$\,$km$\,$s$^{-1}$ (blue contours), overlapped on the mass surface
density map of \citet[][gray scale]{but09}. The contour levels of the SiO emission,
are 10 (2$\sigma$; dashed contour), 15, 20, 30 and 40$\,$mK$\,$km$\,$s$^{-1}$.
For the mass surface density map, contours are 
0.014 (2$\sigma$), 0.021, 0.035, 0.049, 0.07, 0.105 and 0.14$\,$g$\,$cm$^{-2}$, respectively.
Crosses indicate the cores reported in the cloud by \citet[][]{rath06} and 
\citet{but09}. 
Red open circles, red open triangles and green squares show the location of the 
8$\,$$\mu$m sources, 24$\,$$\mu$m sources 
\citep{car09} and 4.5$\,$$\mu$m extended emission \citep{cham09} 
detected in G035.39-00.33, respectively. 
The marker sizes used for the 8$\,$$\mu$m and 24$\,$$\mu$m sources, 
have been scaled by the source flux. Red dashed lines show the 
directions of the $P-V$ diagrams of Figure$\,$\ref{f2}. Red numbers
indicate the distance (in arcseconds) along the cuts made for these diagrams.
The SiO $J$=2$\rightarrow$1 beam size is plotted at the upper right corner. {\it Left 
and right panels}: Sample of spectra of SiO $J$=2$\rightarrow$1 and 
$J$=3$\rightarrow$2, and of CO, $^{13}$CO and C$^{18}$O $J$=2$\rightarrow$1, 
measured toward several positions across the IRDC. The CO, $^{13}$CO,
and C$^{18}$O spectra were obtained by averaging the OTF data within the 
28$''$-beam of the SiO $J$=2$\rightarrow$1 observations. 
The vertical dotted lines show the velocities of the 
CO filaments, as calculated from the C$^{18}$O $J$=2$\rightarrow$1 emission 
(Table$\,$\ref{tab1} and Section$\,$\ref{co}).}
\label{f1}
\end{center}
\end{figure*}

\begin{table}
 \centering
 \begin{minipage}{85mm}
  \caption{Observed molecular transitions and line frequencies, 
  telescope beam sizes and beam efficiencies for the EMIR and HERA receivers 
at the IRAM 30$\,$m telescope.}
  \begin{tabular}{lccc}
  \hline
Transition &  Frequency (MHz) & Beam size ('') & Beam Eff. \\ \hline 
SiO $J$=2$\rightarrow$1 & 86846.96 & 28 & 0.81 \\
SiO $J$=3$\rightarrow$2 & 130268.61 & 19 & 0.74 \\
CO $J$=2$\rightarrow$1 & 230538.00 & 11 & 0.63 \\
$^{13}$CO $J$=2$\rightarrow$1 & 220398.68 & 11 & 0.52 \\
C$^{18}$O $J$=2$\rightarrow$1 & 219560.36 & 11 & 0.52 \\ \hline
\label{tab2}
\end{tabular}
\end{minipage}
\end{table}

\begin{table*}
 \centering
 \begin{minipage}{140mm}
  \caption{Observed parameters of the SiO $J$=2$\rightarrow$1 and $J$=3$\rightarrow$2
lines, and of the $J$=2$\rightarrow$1 emission of C$^{18}$O, toward
several offsets in IRDC G039.35-00.33.}
  \begin{tabular}{lccccccccc}
  \hline
Molecule &  \multicolumn{9}{c}{OFFSETS}\\ 
& \multicolumn{3}{c}{(0,20)} & \multicolumn{3}{c}{(10,10)} & 
\multicolumn{3}{c}{(30,-30)} \\ \cline{2-10}
& $v_{LSR}$ & $\Delta v$ & T$_A^*$ & $v_{LSR}$ & $\Delta v$ & T$_A^*$ & 
$v_{LSR}$ & $\Delta v$ & T$_A^*$ \\
& (km$\,$s$^{-1}$) & (km$\,$s$^{-1}$) & (K) & (km$\,$s$^{-1}$) & 
(km$\,$s$^{-1}$) & (K) & (km$\,$s$^{-1}$) & (km$\,$s$^{-1}$) & (K) \\ \hline
SiO(2$\rightarrow$1) & $\sim$44.0 & $\ldots$ & $\leq$0.03 & 44.73(8) & 0.7(2) & 0.07(1) & 
42.9(4) & 7.0(8) & 0.06(1) \\
    & 45.56(6)& 1.3(2) & 0.13(1) & 45.7(1) & 0.8(3) & 0.06(1) & 44.73(7) & 1.9(2) 
& 0.12(1) \\
    & 47.30(8)& 1.0(2) & 0.07(1) & $\sim$46.8(2) & $\ldots$ & $\leq$0.04 & $\sim$46.4 & $\ldots$ 
& $\leq$0.04 \\
SiO(3$\rightarrow$2) & $\sim$43 & $\ldots$ & $\leq$0.04 & $\ldots$ & $\ldots$ 
& $\ldots$ & 42.9(3) & 3(1) & 0.06(2) \\
                     & 45.5(1)& 1.7(4) & 0.07(1) & $\ldots$ & $\ldots$ & 
$\ldots$ & 44.9(2) & 2.3(6) & 0.08(2) \\
                     & 47.8(2) &  0.8(3) & 0.05(1) & $\ldots$ & $\ldots$ & 
$\ldots$ & $\sim$46.4 & $\ldots$ & $\leq$0.05 \\
C$^{18}$O(2$\rightarrow$1) & 44.0(2) & 2.3(3) & 0.34(3) & 44.500(1) & 3.0(4) & 
0.25(4) & 43.1(2) & 1.7(4) & 0.08(3) \\
                           & 45.41(1)& 1.08(4)& 1.30(3) & 45.38(1) & 1.07(3) & 
1.36(4) & 45.21(2) & 1.30(3) & 0.87(3) \\
                           & 46.77(2)& 1.26(5)& 0.78(3) & 46.78(8) & 1.4(1) & 
0.36(4) & 46.4(1) & 1.4(3) & 0.09(3) \\ \hline
& \multicolumn{3}{c}{(-10,70)} & \multicolumn{3}{c}{(-37,37)} & 
\multicolumn{3}{c}{(0,-80)} \\ \cline{2-10}
SiO(2$\rightarrow$1) & $\sim$44.9 & $\ldots$ & $\leq$0.018 & $\sim$43.9 & 
$\ldots$ & $\leq$0.03 & 43.2(3) & 6.2(3) & 0.04(1) \\
                     & $\sim$45.9 & $\ldots$ & $\leq$0.018 & 45.3(1) & 
0.8(3) & 0.04(1) & 45.6(3) & 2.9(3) & 0.11(1) \\
                     & 46.5(2) & 1.5(4) & 0.021(6) & $\sim$46.7 & $\ldots$ & 
$\leq$0.03 & 49.1(3) & 3.5(3) & 0.06(1) \\
SiO(3$\rightarrow$2) & $\sim$44.9 & $\ldots$ & $\leq$0.04 & $\sim$43.9 & 
$\ldots$ & $\leq$0.05 & $\sim$43.2 & $\ldots$ & $\leq$0.07 \\
                     & $\sim$45.9 & $\ldots$ & $\leq$0.04 & $\sim$45.1 & 
$\ldots$ & $\leq$0.05 & 45.3(1) & 0.9(3) & 0.07(2) \\
                     & $\sim$46.9 & $\ldots$ & $\leq$0.04 & $\sim$46.7 & 
$\ldots$ & $\leq$0.05 & 47.5(5) & 4.8(9) & 0.11(2) \\ 

C$^{18}$O(2$\rightarrow$1) & 44.9(1) & 2.2(2) & 0.57(3) & 43.9(1) & 1.1(1) &
0.69(3) & 44.097(8) & 2.4(4) & 0.13(3) \\   
                          & 45.89(3) & 0.8(1) & 0.60(3) & 45.1(1) & 1.2(1) &
0.33(3) & 45.07(2) & 1.22(3) & 1.07(3) \\ 
                          & 46.92(5) & 1.34(9) & 0.68(3) & 46.7(1) & 1.7(1) &
0.40(3) & 45.92 (3)& 1.04(6) & 0.56(3) \\ \hline
\label{tab1}
\end{tabular}
\end{minipage}
\end{table*}

The $J$=2$\rightarrow$1 lines of SiO, CO, $^{13}$CO and C$^{18}$O, were mapped 
with the IRAM (Instituto de Radioastronom\'{\i}a Milim\'etrica) 
30$\,$m telescope at Pico Veleta (Spain) over an area of 2$'$$\times$4$'$ 
toward G035.39-00.33. These observations were carried out in August 2008, and 
in January and February 2009. The large-scale molecular images were obtained 
in the On-The-Fly (OTF) mode using the offsets (1830$''$,658$''$) for SiO, $^{13}$CO 
and C$^{18}$O, and (4995$''$,2828$''$) for CO, as off-positions. The  
central coordinates of the map were $\alpha$(J2000)=18$^h$57$^m$08$^s$, 
$\delta$(J2000)=02$^\circ$10$'$30$''$ (l=35.517$^\circ$, b=$\,$-0.274$^\circ$). The SiO 
$J$=2$\rightarrow$1 emission was mapped with the old SIS receivers, 
while the HERA multi-beam receivers simultaneously observed the 
$J$=2$\rightarrow$1 transitions of $^{13}$CO and C$^{18}$O. The CO 
$J$=2$\rightarrow$1 emission was mapped with the new generation EMIR receivers. 
In addition, we carried out single-pointing observations of the 
SiO $J$=2$\rightarrow$1 and $J$=3$\rightarrow$2 emission with EMIR toward the offsets 
(0,20), (30,-30), (0,-80), (-10,70) and (-37,37). The former 
three positions correspond to the brightest SiO emission peaks observed
toward the IRDC (Section$\,$\ref{nms}). The latter two offsets show the regions
where we have detected narrow SiO lines (Section$\,$\ref{narrow}).  
All receivers were 
tuned to single sideband (SSB) with rejections of $\geq$10$\,$dB. 
The beam sizes were 28$''$ at 90$\,$GHz for the 
SiO $J$=2$\rightarrow$1 line, 19$''$ at 130$\,$GHz for the SiO $J$=3$\rightarrow$2
emission, and 11$''$ at 230$\,$GHz for the CO, $^{13}$CO and C$^{18}$O 
$J$=2$\rightarrow$1 lines. The VESPA spectrometer 
provided spectral resolutions of 
40 and 80$\,$kHz, which correspond to velocity resolutions of $\sim$0.14 and 
0.1$\,$km$\,$s$^{-1}$ at 90 and 230$\,$GHz, respectively. Saturn was used
to calculate the focus, and pointing was checked every two hours on 
G34.3+0.2. Typical system temperatures ranged from 100 to 300$\,$K. All
intensities were calibrated in units of antenna temperature, $T_A^*$. To convert
these intensities into units of main-beam temperature, $T_{mb}$, we have used 
efficiencies of 0.81, 0.74 and 0.63 for the EMIR data at $\sim$90, 130 and 230$\,$GHz, 
and of 0.52 for the HERA data at $\sim$230$\,$GHz. All this information is
summarized in Table$\,$\ref{tab2}.

Figure$\,$\ref{f1} (central panel) 
presents the high-sensitivity map of the SiO $J$=2$\rightarrow$1 emission
integrated from 36 to 51.5$\,$km$\,$s$^{-1}$ (blue contours), and 
superimposed on the mass surface density map, with an angular
resolution of 2$''$, reported by 
\citet[][]{but09}. 
The SiO $J$=2$\rightarrow$1 map has been obtained by averaging the
OTF dumps in the SiO beam of 28$''$, and by using a Nyquist-sampled
grid with a pixel size of 14$''$. 
The 2$\sigma$ intensity level of SiO is shown in dashed contours. 
The location of the massive cores \citep[crosses;][]{rath06,but09}, 24$\,$$\mu$m sources 
\citep[red open triangles; extracted from MIPSGAL images;][]{car09}, 
8$\,$$\mu$m sources (red open circles), and 
4.5$\,$$\mu$m extended emission \citep[green squares; called {\it green fuzzies}
in][]{cham09} in this IRDC, are also shown. The flux lower limits of the 
8$\,$$\mu$m and 24$\,$$\mu$m sources reported in Figure$\,$\ref{f1}, 
are $\geq$3.5$\,$mJy and $\geq$2$\,$mJy, respectively. We note 
that the cavity-like structures seen around the 8$\,$$\mu$m sources, 
are produced by the fact that the extinction mapping 
technique of \citet{but09} cannot be applied in the vicinity of 
IR-bright sources.

From Figure$\,$\ref{f1}, we find that the 
SiO $J$=2$\rightarrow$1 emission is widespread 
across the filament with a spatial extent of $\geq$150$''$$\times$50$''$. This
corresponds to $\geq$2.1$\,$pc$\times$0.7$\,$pc at a distance of 
$\sim$2.9$\,$kpc \citep[][]{rath06}. This size should be considered as a lower limit 
since SiO is also detected toward the north and northwest of Core MM7/H6
(see Section$\,$\ref{narrow}), and 
extends off to the south of the imaged area.    
The SiO map shows three bright condensations (marked N, E, S) surrounded  
by weaker, more extended emission, that covers much of the IRDC  
filament.

In Figure$\,$\ref{f1}, we also report a sample of spectra of 
SiO $J$=2$\rightarrow$1 and $J$=3$\rightarrow$2, and of the CO, $^{13}$CO 
and C$^{18}$O $J$=2$\rightarrow$1 lines, measured toward several 
positions in the IRDC. Since single-pointing observations
of SiO $J$=2$\rightarrow$1 and $J$=3$\rightarrow$2 were not carried out
toward (10,10), the SiO $J$=2$\rightarrow$1 spectrum reported in 
Figure$\,$\ref{f1} was obtained by averaging the 
OTF data within the 28$''$-beam of the IRAM 30$\,$m telescope at 
$\sim$90$\,$GHz. The spectra of CO and of its isotopologues
shown in Figure$\,$\ref{f1}, 
have also been obtained following this procedure. 
From the SiO spectra, we find that
SiO shows a wide variety of line profiles, 
from broad emission with red- and/or blue-shifted line wings [see offsets
(0,20), (30,-30) and (0,-80)], to narrow lines 
peaking at the ambient cloud velocity  
$v_{LSR}$$\sim$45$\,$km$\,$s$^{-1}$ [offsets (10,10), (-10,70) and (-37,37)]. 
The CO, $^{13}$CO and C$^{18}$O lines 
show three different velocity components centred at 
$v_{LSR}$$\sim$44, 45 and 47$\,$km$\,$s$^{-1}$ (see Section$\,$\ref{co}).

In Table$\,$\ref{tab1}, we report the observed parameters (central radial
velocity, $v_{LSR}$, linewidth, $\Delta v$, and peak intensity, $T_A^*$) 
of the different velocity components measured in SiO $J$=2$\rightarrow$1, 
SiO $J$=3$\rightarrow$2 and C$^{18}$O $J$=2$\rightarrow$1 
(representative of the low-density CO gas) toward (0,20), 
(10,10), (30,-30), (-10,70), (-37,37) and (0,-80). These parameters were 
obtained by fitting the molecular line emission with three gaussian line profiles
simultaneously. Although this method works better for the C$^{18}$O 
$J$=2$\rightarrow$1 emission, the gaussian linewidths derived from the SiO 
$J$=2$\rightarrow$1 line profiles provide a rough estimate of the velocity 
range extent and terminal velocities of the shocked SiO emission. 
In Table$\,$\ref{tab1}, the errors in $v_{LSR}$ and $\Delta v$ 
correspond to those obtained from the multi-component gaussian fit. For the errors in the 
peak intensities, $T_A^*$, we consider the noise r.m.s (1$\sigma_{rms}$) 
of the spectra 
in Figure$\,$\ref{f1}, and for the upper limits, the 3$\sigma_{rms}$ noise level of this emission. 

\begin{figure}
\includegraphics[angle=270,width=0.47\textwidth]{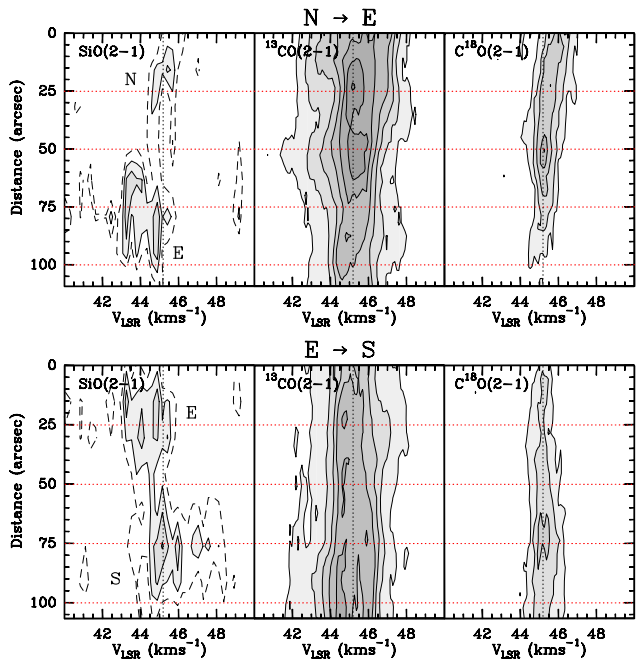}
\caption{$P-V$ diagrams of the SiO, $^{13}$CO and C$^{18}$O 
$J$=2$\rightarrow$1 emission (in units of T$_A^*$), obtained along the 
lines linking condensations N, E and S (see red dashed lines and red 
numbers in Figure$\,$\ref{f1}). The distance (in arcseconds)
shown in the $y$-axis corresponds to that measured along these lines 
from north to south (see also horizontal red dotted lines
in this Figure). The contour levels 
for SiO $J$=2$\rightarrow$1 are 60 (2$\sigma$; dashed contours), 90, 
120 and 150$\,$mK. For $^{13}$CO and C$^{18}$O, the first contour and 
level step are 0.48 (3$\sigma$) and 0.8$\,$K for 
$^{13}$CO $J$=2$\rightarrow$1, and 0.48 (3$\sigma$) and 0.48$\,$K for 
C$^{18}$O $J$=2$\rightarrow$1. While the beam size of the 
SiO $J$=2$\rightarrow$1 data is 28$''$, the angular resolution 
used in the $P-V$ diagrams of the $^{13}$CO and C$^{18}$O 
$J$=2$\rightarrow$1 lines, is 11$''$.}
\label{f2}
\end{figure}

\subsection{The broad SiO line emission toward condensations N, E and S}
\label{nms}

Figure $\,$\ref{f1} shows that the brightest SiO $J$=2$\rightarrow$1
and $J$=3$\rightarrow$2 emission in the IRDC
arises from three major condensations toward (0,20) [N], (30,-30) [E] 
and (0,-80) [S]. The N and S condensations peak at the densest cores 
reported in the filament \citep[MM7/H6 and MM6/H5;][]{rath06,but09}, 
and harbor not only several 8$\,$$\mu$m and 24$\,$$\mu$m 
sources, but 
4.5$\,$$\mu$m extended emission \citep[likely related to H$_2$ 
shocked gas;][]{nor04}. This suggests that these condensations  
are active sites of star formation. 

Condensation E, however, is located east of the IRDC and 
slightly off the high extinction region (see Figure$\,$\ref{f1}). 
Although small amounts of gas are present, 
no local maxima is seen toward E 
in either extinction 
\citep[][]{but09}, or in other high-density molecular 
tracers such as N$_2$H$^{+}$ or H$^{13}$CO$^{+}$ 
(Caselli et al. 2010, in preparation). Like N and S, 
several 8$\,$$\mu$m sources are detected toward this 
condensation, but only one is seen at 24$\,$$\mu$m
and this is about 15$''$ north of  
the peak SiO position (see Figure$\,$\ref{f1}). The Spitzer IRAC (3-8$\,$$\mu$m) 
and MIPS 24$\,$$\mu$m fluxes of this source are consistent 
with a protostellar model with a luminosity of 
$\sim$2$\times$10$^4$$\,$L$_\odot$, negligible circumstellar material, and  
a foreground extinction of A$_V$=15$\,$mag. This corresponds to a 15$\,$M$_\odot$ 
star on the zero age main sequence (ZAMS). This is the best fit model returned 
from the SED fitting program of \citet{rob07}. However, this result is not unique
and the determination of protostellar properties from a relatively poorly 
constrained SED is quite uncertain. For example, the observed luminosity across the IRAC bands  
is only $\sim$20$\,$L$_\odot$, and the above estimate of a much higher  
luminosity relies on the reality of the large foreground extinction.
The source in condensation E could be responsible for the SiO broad feature observed  
toward this condensation. However, one would not expect strong outflow  
activity from such an evolved star. Alternatively, 
this source could be of much lower luminosity and 
mass and has more active accretion and outflow activity. 
There may also be other low-luminosity protostars in the vicinity, although
the lack of high enough angular resolution in our SiO images prevents to 
establish if this is the origin of the broad SiO emission in this condensation (see below).

In Figure$\,$\ref{f2}, we show the $P-V$ diagrams of the SiO 
$J$=2$\rightarrow$1 emission observed between the N and E, and the 
E and S condensations, for the velocity range from
40 to 50$\,$km$\,$s$^{-1}$. Outside this velocity range, the
emission of SiO $J$=2$\rightarrow$1 is below the 2$\sigma$ rms level
in the spectra. From Figure$\,$\ref{f2}, we find that
the typical SiO line profiles toward condensations N, E and S 
have a central component peaking at 
$v_{LSR}$$\sim$45$\,$km$\,$s$^{-1}$, with
broader line wing emission. The linewidths of the central 
component are $\sim$1-3$\,$km$\,$s$^{-1}$, and those of the broad 
SiO emission are $\sim$4-7$\,$km$\,$s$^{-1}$ (Table$\,$\ref{tab1}).
The blue- and red-shifted terminal velocities are  
$\sim$40 and 50$\,$km$\,$s$^{-1}$, respectively (i.e. $\pm$5$\,$km$\,$s$^{-1}$
with respect to the central velocity $v_{LSR}$$\sim$45$\,$km$\,$s$^{-1}$; 
Figure$\,$\ref{f1}). While the broad SiO emission toward 
condensation E is blue-shifted, the SiO linewings toward N and S 
mainly appear at red-shifted velocities. Broad line profiles 
have previously been reported in SiO and other molecular species 
such as H$_2$CO and CH$_3$OH toward several samples of IRDCs
\citep[see][]{car98,beu07,leu07,sak08}, and 
are believed to trace material associated with molecular outflows. 

The SiO $P-V$ diagram of Figure$\,$\ref{f2} also shows 
that condensations N, E and S have several local maxima centered
at different radial velocities. Due to the low angular
resolution of our SiO observations, it is currently impossible to  
determine whether these maxima are produced by error fluctuations
in the SiO line temperature, or whether they are associated with shocked 
gas high-velocity {\it bullets}, or with different low-mass protostars
(see Section$\,$\ref{ori}).

The $P-V$ diagrams of $^{13}$CO $J$=2$\rightarrow$1 
(Figure$\,$\ref{f2}) show that the terminal velocities of the
$^{13}$CO lines are, in general, similar to those
measured for SiO toward the condensations N, E and S. 
The C$^{18}$O $J$=2$\rightarrow$1
emission does not reveal any significant broad line wing emission
since it is mainly associated with the high-density gas seen in
extinction toward G035.39-00.33 (see Section$\,$\ref{co}). The
correlation between the CO molecular gas and the mass 
surface density map of \citet{but09}, will be analysed in detail 
in the near future (Hernandez et al. 2010, in preparation).

\subsection{The extended and narrow SiO components}
\label{narrow}

\begin{figure*}
\includegraphics[angle=270,width=1.0\textwidth]{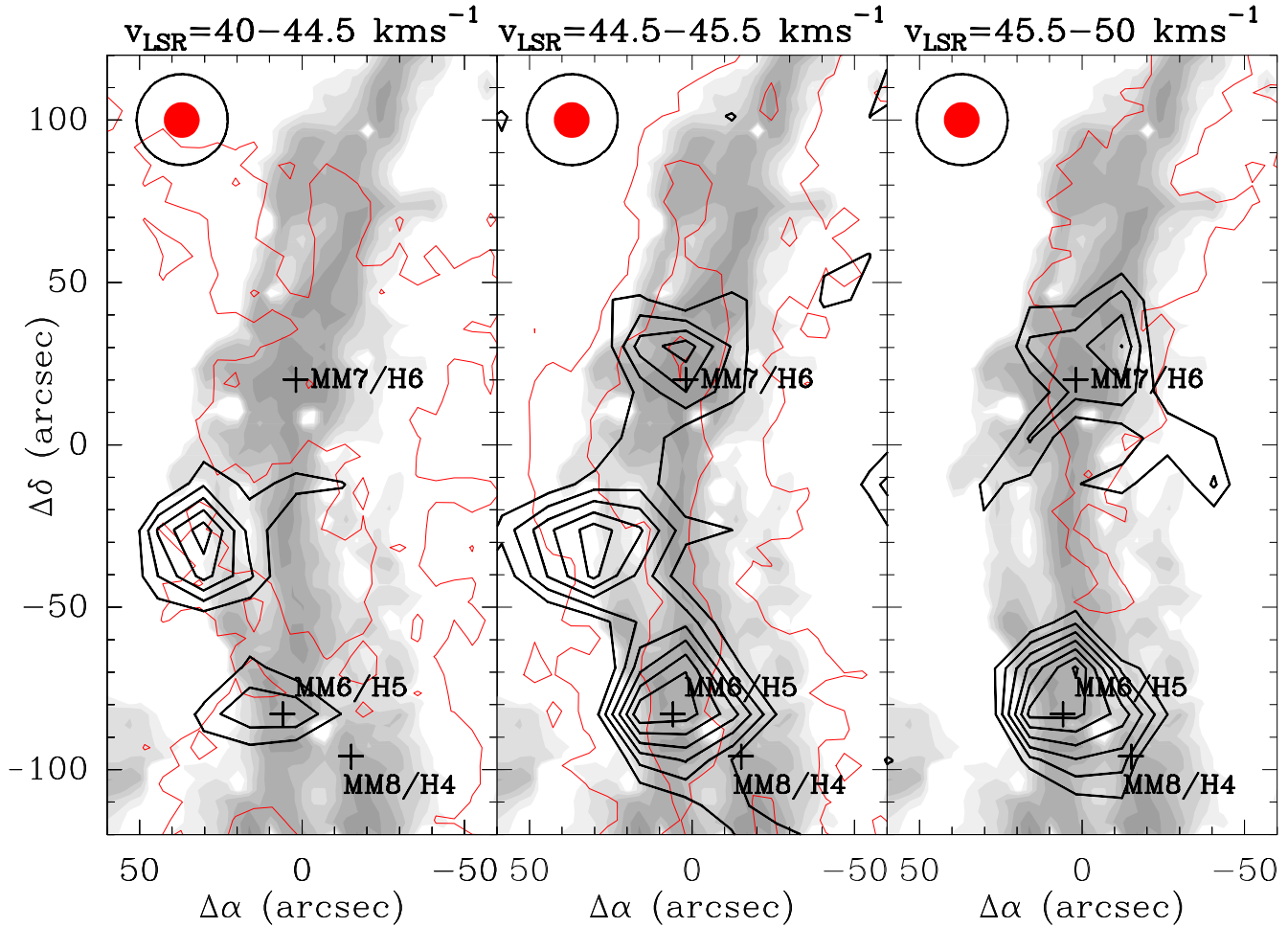}
\caption{Integrated intensity maps (in units of T$_A^*$$\,$km$\,$s$^{-1}$) 
of the SiO $J$=2$\rightarrow$1 
(black contours) and C$^{18}$O $J$=2$\rightarrow$1 lines 
(red contours) measured toward G035.39-00.33 
from 40 to 44.5$\,$km$\,$s$^{-1}$ (blue-shifted gas; left panel), 
44.5 to 45.5$\,$km$\,$s$^{-1}$ (ambient gas; central panel),
and 45.5 to 50$\,$km$\,$s$^{-1}$ (red-shifted gas; right panel). 
The mass surface density map \citep{but09} is shown in greyscale (contour levels as in
Figure$\,$\ref{f1}). The contour levels of the SiO 
$J$=2$\rightarrow$1 maps are 24 (3$\sigma$), 32, 40, 48 and 
56$\,$mK$\,$km$\,$s$^{-1}$ for the blue-shifted
emission, 51 (3$\sigma$), 68, 85, 102, 119 and 
136$\,$mK$\,$km$\,$s$^{-1}$ for the ambient velocity range, and 27 
(3$\sigma$), 36, 45, 54, 63, 72 and 81$\,$mK$\,$km$\,$s$^{-1}$ 
for the red-shifted gas. 
For clarity, we only plot the contours at the 5$\sigma$, 15$\sigma$ and 25$\sigma$ levels in the C$^{18}$O maps, 
with first contours at 150, 350 and 
200$\,$mK$\,$km$\,$s$^{-1}$ for the blue-shifted (left), 
ambient (centre), and red-shifted (right) velocity
ranges, respectively. 
Beam sizes of the SiO $J$=2$\rightarrow$1 (black circle) 
and C$^{18}$O $J$=2$\rightarrow$1
observations (filled red circle) are indicated in the upper left corner.}
\label{f3}
\end{figure*}

In addition to the broad SiO condensation N, E and S, 
the $P-V$ diagrams of Figure$\,$\ref{f2} show that very narrow SiO emission [with 
linewidths of $\leq$1$\,$km$\,$s$^{-1}$; offset (10,10) in Figure$\,$\ref{f1}] 
arises from regions linking these condensations. This is
more clearly seen in Figure$\,$\ref{f3} (central panel), where 
the narrow SiO lines arising from ambient gas 
(at $v_{LSR}$$\sim$45$\,$km$\,$s$^{-1}$), form 
a large-scale and extended {\it ridge} that follows the filament. We note that the 
SiO emission associated with the ridge shows
narrower line profiles than C$^{18}$O $J$=2$\rightarrow$1 
[0.8$\,$km$\,$s$^{-1}$ vs. $\sim$1-3$\,$km$\,$s$^{-1}$; offset (10,10) in
Table$\,$\ref{tab1}]. The peak intensity of narrow SiO toward this position is relatively 
weak (0.06$\,$K; Table$\,$\ref{tab1}) and is at the $\sim$5$\sigma$ level (the $\sigma_{rms}$ 
of the SiO $J$=2$\rightarrow$1 spectrum is 0.012$\,$K; Figure$\,$\ref{f1}).  

\begin{figure}
\includegraphics[angle=270,width=0.47\textwidth]{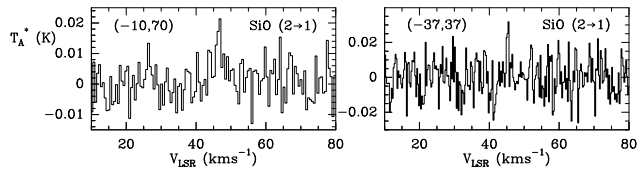}
\caption{High-sensitivity SiO $J$=2$\rightarrow$1 spectra measured with EMIR 
toward ($-$10,70) and ($-$37,37). The rms is 6$\,$mK for the ($-$10,70) 
spectrum and 10$\,$mK for the ($-$37,37) observation.}
\label{f4}
\end{figure}

The high-sensitivity single-pointing SiO spectra obtained with EMIR 
toward (-10,70) and (-37,37), also reveal that the
narrow SiO component spreads north and northwest of Core MM7/H6 (Figure$\,$\ref{f4}). 
The SiO data toward (-10,70) has been smoothed to a velocity
resolution of 0.53$\,$km$\,$s$^{-1}$ to improve the signal-to-noise 
ratio of the spectrum. For the (-37,37) offset, however, we keep
a velocity resolution of 0.26$\,$km$\,$s$^{-1}$, because the SiO line 
emission toward this position is a factor of 2 narrower than that reported
toward (-10,70) (i.e. 0.8$\,$km$\,$s$^{-1}$ vs. 1.5$\,$km$\,$s$^{-1}$; 
see Table$\,$\ref{tab1}).  The narrow SiO lines are
very faint and have integrated line intensities of
0.038$\pm$0.005$\,$K$\,$km$\,$s$^{-1}$ and 0.030$\pm$0.005$\,$K$\,$km$\,$s$^{-1}$ toward (-10,70) and (-37,37), respectively. 
Since the SiO lines reported by \citet{beu07} in a sample of IRDCs have significantly  
larger linewidths ($\geq$2.5$\,$km$\,$s$^{-1}$) than those observed toward 
the ridge, toward (-10,70) or toward (-37,37), these lines 
are the narrowest features detected so far in a high-mass star forming region. 
The SiO $J$=3$\rightarrow$2 lines are not detected toward positions (-10,70) and (-37,37). 

From the weak intensity and spatial distribution of narrow SiO,
one could think that this emission could be due to some line emission 
contribution, within the large 30$\,$m beam of our observations 
($\sim$28$''$), arising from condensations N, E and S. 
However, the narrow SiO lines measured toward (10,10), (-10,70)
and (-37,37) have line profiles different from those observed
toward N, E, and S (see Figure$\,$\ref{f1}), and their central radial 
velocities differ from those found toward these
condensations. This is particularly clear toward (-10,70),
where the line peak velocity of narrow SiO is red-shifted by 
$\sim$1$\,$km$\,$s$^{-1}$ with respect to that derived toward N. 
Therefore, the narrow SiO lines detected in G035.39-00.33 
trace different molecular material from that seen in the SiO 
condensations N, E and S.

Finally, in Figure$\,$\ref{f1}, we note that the narrow SiO emission toward (-10,70)
and (-37,37) lies below the 2$\sigma$ contour level of the integrated intensity 
SiO map. This is due to the fact that narrow SiO lines are diluted in the 
broad velocity range considered to create the map. 
In Figure$\,$\ref{f3} (central panel), the narrow SiO emission 
detected toward (-10,70) with the high-sensitivity EMIR receivers 
(Figure$\,$\ref{f4}), also lies below the 3$\sigma$ noise level of
 the SiO map at ambient velocities, because the observations with the 
old SIS receivers were not sensitive enough to detect such faint 
emission (see above). The narrow SiO lines toward (-10,70) are of 
particular interest because they do not show any clear association 
with a 8$\,$$\mu$m or a 24$\,$$\mu$m source.  
The narrow SiO emission detected toward (-37,37) is likely
associated with the 
faint SiO condensation (intensity level of $\sim$3$\sigma$) located 
at (-40,50) in the central panel of Figure$\,$\ref{f3}.

\subsection{The CO filaments in G035.39-00.33}
\label{co}

The general kinematics of the low-density CO gas toward G035.39-00.33, 
as traced by C$^{18}$O $J$=2$\rightarrow$1, are also shown in Figure$\,$\ref{f3}. 
The CO gas is distributed along three different filaments with radial velocities  
$v_{LSR}$=44.1, 45.3 and 46.6$\,$km$\,$s$^{-1}$. These values 
correspond to the averaged central radial velocities derived for every filament
from the C$^{18}$O $J$=2$\rightarrow$1 gaussian fit parameters shown 
in Table$\,$\ref{tab1}. While the central and brightest 
filament at $v_{LSR}$$\sim$45$\,$km$\,$s$^{-1}$ bends east tightly following 
the densest material within the IRDC, the blue-shifted filament with 
$v_{LSR}$$\sim$44$\,$km$\,$s$^{-1}$ intersects the 
former one in an arc-like structure pointing west (i.e. with the centre
of curvature lying to the east of the IRDC). 
The two intersecting regions are coincident with the highest density cores reported 
in the IRDC, MM7/H6, MM6/H5 and MM8/H4 \citep[][]{rath06,but09}. 
Although not as clear as for the 
blue-shifted CO component, the red-shifted filament with 
$v_{LSR}$$\sim$47$\,$km$\,$s$^{-1}$ seems to further bend east with 
respect to the central one, following the trend already shown by 
the latter filament. The morphology of these filaments resembles
the molecular structures predicted by flow-driven, shock-induced and 
magnetically-regulated models of cloud formation 
\citep[see e.g.][]{fie00a,fie00b,van07,hen08,heit09}. 
As discussed in Section$\,$\ref{ori}, in these scenarios
the high-density 
filament seen in extinction would have been formed in 
a cloud-cloud collision, with the low-density CO filaments 
still accreting material onto it. 
The kinematics and excitation of the low-density CO gas 
will be described in a forthcoming paper (Jim\'enez-Serra et al. 2010, in preparation).

\section{Excitation and fractional abundances of SiO}
\label{exc}

Except for the (10,10) position, where SiO $J$=3$\rightarrow$2 single-dish observations
were not carried out (see Section$\,$\ref{res}), we can derive the H$_2$ densities of the gas, 
$n$(H$_2$), and the excitation temperatures, T$_{ex}$, 
and column densities of SiO, $N$(SiO), from the SiO $J$=3$\rightarrow$2/$J$=2$\rightarrow$1
intensity ratio and by using the Large Velocity Gradient 
(LVG) approximation. As shown in Section$\,$\ref{val}, the validity
of this approximation is provided by the fact that the SiO line emission 
toward G035.39-00.33 is optically thin.
We assume that the SiO $J$=2$\rightarrow$1 and 
$J$=3$\rightarrow$2 lines trace the same material.

\subsection{LVG modelling and input parameters}
\label{lvg}

In our LVG calculations, we have considered the first 15 rotational levels of SiO, and 
used the collisional coefficients of SiO with H$_2$ derived by \citet{tur92}
for temperatures up to $T$=300$\,$K. These collisional
coefficients are well suited for our case of study, because the 
kinetic temperatures assumed for the SiO shocked
gas in the IRDC are $T$$\ll$300$\,$K. As input parameters, we have considered 
the CMB (Cosmic Microwave Background) temperature (i.e. $T_{bg}$=2.7$\,$K), 
the linewidths of the SiO emission, and the kinetic temperature of the gas, 
$T_{kin}$. The brightness temperature, $T_B$, excitation temperature, 
$T_{ex}$, and optical depth, $\tau$(SiO), of every rotational line transition
of SiO, are then calculated for a grid of models with different H$_2$ 
(volumetric) gas densities and column densities of SiO. We note 
that $T_{ex}$ may 
significantly differ from $T_{kin}$, since the excitation conditions
of the SiO shocked gas could be far away from the LTE (Local Thermodynamic
Equilibrium).

Three different velocity regimes have been considered for 
the SiO emission observed in G035.39-00.33: the ambient component, which 
ranges from 44.5 to 45.5$\,$km$\,$s$^{-1}$; the moderate shocked gas, with 
42.5$\,$km$\,$s$^{-1}$$\leq$$v_{LSR}$$<$44.5$\,$km$\,$s$^{-1}$ or 
45.5$\,$km$\,$s$^{-1}$$<$$v_{LSR}$$\leq$47.5$\,$km$\,$s$^{-1}$; 
and the high-velocity regime, with $v_{LSR}$$<$42.5$\,$km$\,$s$^{-1}$ or 
$v_{LSR}$$>$47.5$\,$km$\,$s$^{-1}$. Within every velocity regime, 
we perform the LVG calculations for velocity bins of 
1$\,$km$\,$s$^{-1}$. We assume that $T_{kin}$=15$\,$K for the ambient 
component \citep[similar to those derived in a sample of 
IRDCs;][]{pill07}, and $T_{kin}$=25$\,$K 
and $T_{kin}$=45$\,$K for the moderate and high-velocity 
SiO gas, respectively. The latter two temperatures are consistent with those
found in the shocked gas of the L1448-mm outflow \citep[][]{jim05}.

We would like to stress that the selection of these temperatures
are expected not to be crucial in our calculations of $N$(SiO), 
since the excitation of the SiO rotational lines with $J_{up}$$<$5 does not strongly 
depend on $T_{kin}$ but on the H$_2$ density of the gas \citep{nis07}. 
Indeed, if we increase $T_{kin}$ from 15$\,$K to 50$\,$K for the
narrow SiO component, and from 25-45$\,$K to 300$\,$K for the moderate and
high-velocity SiO shocked gas, the derived $N$(SiO)
change by less than a factor of 2.

The brightness temperatures, $T_B$, derived with the LVG model 
for the SiO $J$=2$\rightarrow$1 and $J$=3$\rightarrow$2 lines,
are finally compared with those observed toward G035.39-00.33, in 
units of $T_{mb}$. We assume that the SiO emission is uniformly distributed 
and that the beam-filling factor of our SiO observations is $\sim$1 (only in this
case $T_B$$\sim$$T_{mb}$). This is justified by the fact that the
SiO emission is extended in the IRDC.

To derive the SiO fractional abundances, the H$_2$ column
densities were estimated from C$^{18}$O for the ambient gas, from $^{13}$CO for the
moderate velocity regime, and from CO for the high-velocity gas. 
In contrast with the mass surface density map of \citet{but09}, 
which gives an averaged value of the H$_2$ column density of 
the gas toward a certain position, CO and its isotopologues 
provide estimates of the H$_2$ column density as a function 
of velocity within the shock.
We assume isotopic ratios $^{12}$C/$^{13}$C=53 and 
$^{16}$O/$^{18}$O=327 \citep{wil94}, and a CO
fractional abundance of 2$\times$10$^{-4}$ across the IRDC. The uncertainty in 
the CO abundance is about a factor of two, considering its variations in 
different molecular cloud complexes \citep[e.g.][]{frer82} as well as within 
the same complex \citep{pin08}. The value adopted in this study is close 
to that directly measured towards another high mass star forming region 
\citep[NGC2024;][]{lacy94}, which better represents the properties of IRDCs.

\subsection{LVG results}
\label{lvgres}

\begin{figure}
\includegraphics[angle=270,width=0.5\textwidth]{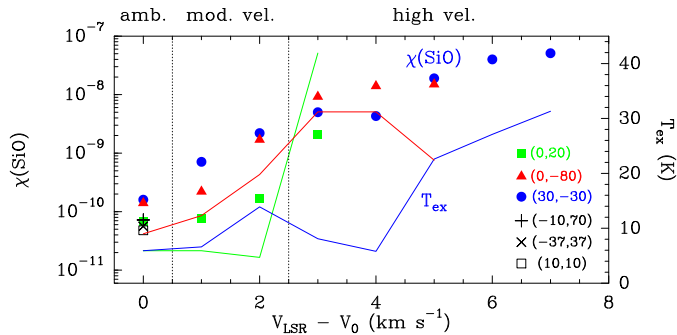}
\caption{Excitation temperatures, T$_{ex}$ (color lines), and SiO fractional abundances, 
$\chi$(SiO) (symbols), derived for the ambient, moderate and high-velocity 
regimes (vertical dotted lines) observed toward several offsets in 
G035.39-00.33. We consider $v_0$=45.0$\,$km$\,$s$^{-1}$. The errors 
associated with T$_{ex}$ are estimated to range from 15\% to 35\%.}
\label{f5}
\end{figure}

The T$_{ex}$ and SiO fractional abundances, $\chi$(SiO), derived toward  
(-10,70), (-37,37), (0,20), (30,-30) and (0,-80) by means of the LVG approximation, 
are shown in Figure$\,$\ref{f5}. Toward (10,10), $\chi$(SiO) was calculated 
assuming that $T_{ex}$=9$\,$K [i.e. similar to those derived toward (-10,70) and 
(-37,37); see below] and that the SiO emission is optically thin.

For the narrow SiO emission toward ($-$10,70) and ($-$37,37), 
we obtain T$_{ex}$$\sim$9$\,$K, optical depths $\tau$(SiO)$\leq$0.01,
and SiO column densities ranging from 5$\times$10$^{10}$$\,$cm$^{-2}$
to 10$^{11}$$\,$cm$^{-2}$. This implies SiO fractional abundances
$\chi$(SiO)$\sim$6-7$\times$10$^{-11}$ (Figure$\,$\ref{f5}). The derived 
H$_2$ gas densities are $\leq$6$\times$10$^5$$\,$cm$^{-3}$. Toward (10,10),
the derived SiO fractional abundance is $\sim$5$\times$10$^{-11}$.
These abundances are a factor of 10 larger than the upper limits found 
in dark clouds \citep[$\leq$10$^{-12}$;][]{ziu89,req07}, and are similar to those 
measured from narrow SiO toward the molecular 
outflows in the low-mass star forming regions
NGC1333 and L1448-mm \citep{lef98,jim04}. 

Toward the N, E and S condensations, $T_{ex}$ and $\chi$(SiO) tend 
to increase from the ambient 
to the moderate and the high-velocity regimes (Figure$\,$\ref{f5}).
The typical optical depths derived for the SiO emission in 
these condensations are $\tau$(SiO)$\leq$0.06.  
The derived H$_2$ gas densities and SiO column densities 
range from 10$^{5}$$\,$cm$^{-3}$ to 10$^{6}$$\,$cm$^{-3}$, and 
from 5$\times$10$^{10}$$\,$cm$^{-3}$ to 4$\times$10$^{11}$$\,$cm$^{-3}$, 
respectively. For $T_{ex}$, a similar behavior for the  
excitation of the SiO shocked gas has been 
reported toward the L1157-mm and L1448-mm outflows \citep{nis07},
where the SiO $J$=8$\rightarrow$7/$J$=5$\rightarrow$4 ratio is known to
increase as a function of velocity within the shock \citep{nis07}.
In the case of the SiO fractional abundances, 
$\chi$(SiO) is progressively 
enhanced from $\sim$10$^{-10}$ in the ambient gas, to $\sim$10$^{-9}$ in 
the moderate velocity component, and to $\geq$10$^{-8}$ in the 
high-velocity shocked gas. This trend has also been observed toward the L1448-mm 
outflow \citep{jim05}.

Although the LVG model does not provide the errors
associated with T$_{ex}$, we can roughly estimate them from the 
integrated intensity ratio between the SiO $J$=3$\rightarrow$2 and the 
SiO $J$=2$\rightarrow$1 transitions. The errors range from 15\% to 35\%. 
This range corresponds to the change in signal-to-noise ratio going from 
the strong SiO lines at ambient velocities to the fainter SiO emission
at the moderate and high-velocity shock regimes.

\subsection{Validity of the LVG approximation}
\label{val}

As shown in Section$\,$\ref{lvgres}, the typical optical depths 
derived for the SiO emission toward G035.39-00.33, are $\tau$(SiO)$<<$1. 
In molecular outflows, the SiO line emission is expected to be optically 
thin for ratios 
$N$(SiO)/$\Delta v$$<$5$\times$10$^{13}$$\,$cm$^{-2}$$\,$km$^{-1}$$\,$s,
where $N$(SiO) is the derived SiO column density and $\Delta v$, the 
linewidth of the SiO line profiles \citep{nis07}. The derived SiO
column densities, $N$(SiO), toward the IRDC are 
relatively low and range from 5$\times$10$^{10}$$\,$cm$^{-2}$ to 
4$\times$10$^{11}$$\,$cm$^{-2}$ (Section$\,$\ref{lvgres}). Considering
that the velocity bins used in our calculations are 1$\,$km$\,$s$^{-1}$-wide,
the ratio $N$(SiO)/$\Delta v$ is 5$\times$10$^{10}$-4$\times$10$^{11}$$\,$cm$^{-2}$$\,$km$^{-1}$$\,$s, 
i.e. well below the upper limit established by \citet{nis07}. 
Therefore, the use of the LVG approximation in our case is fully justified.

\subsection{Comparison of the SiO fractional abundances with shock modelling}
\label{model}

By using the parametrized model of \citet{jim08}
that mimics the steady-state physical structure of 
a perpendicular C-shock, we find that the 
low SiO fractional abundances of $\sim$10$^{-11}$ associated with the narrow 
SiO emission in G035.39-00.33, could be generated by the sputtering of dust grains 
within C-shocks with $v_s$$\sim$12$\,$km$\,$s$^{-1}$. 
A minimum ion-neutral drift velocity of 
$v_{d}$=$|v_n-v_i|$=5$\,$km$\,$s$^{-1}$ is  
required to sputter SiO abundances of a few 10$^{-11}$ from the icy mantles 
of dust grains, if a fractional abundance of 10$^{-8}$ for SiO is 
present within the mantles \citep{jim08}. This SiO abundance is the same
as that measured in the moderate velocity SiO shocked gas in L1448-mm, 
and believed to trace material recently released from the mantles of dust 
grains \citep{jim05}. If SiO were only present within the cores of dust 
grains, a shock velocity of $v_s$$\sim$30$\,$km$\,$s$^{-1}$, with an 
ion-neutral drift velocity of $v_d$$\sim$18$\,$km$\,$s$^{-1}$,
would be required to reproduce the low SiO fractional abundances of 
few 10$^{-11}$ for narrow SiO. 

The progressive enhancement of the SiO fractional abundances from 
$\sim$10$^{-10}$ to $\geq$10$^{-8}$ toward the N, E and S condensations,
could be explained by the sputtering of the mantles of dust grains 
along the propagation of a C-shock with $v_s$$\sim$30$\,$km$\,$s$^{-1}$, 
as proposed for the L1448-mm outflow \citep[][]{jim08}. However, 
we note that other mechanisms such as shattering or vaporisation in non-dissociative 
J-shocks \citep{gui07,gui09}, could additionally contribute to the 
processing of dust grains in these regions. The inclusion of those mechanisms
in the currently available state-of-the-art shock descriptions, might significantly 
change our understanding of the shocked SiO emission in molecular outflows.

The comparison of synthetic SiO line profiles (as derived 
from radiative transfer modelling) with the SiO line emission observed 
toward the IRDC, lies outside the scope of the present paper. 
The lack of high enough angular resolution in our SiO OTF maps, indeed
prevents to clearly establish the actual morphology and size of the SiO 
emitting regions, which are crucial parameters in the modelling 
of the SiO line profiles. 

\section{On the origin of the parsec-scale SiO emission in G035.39-00.33}
\label{ori}

Theoretical models of flow-driven \citep[][]{hen99,heit06,hen08,heit09},
shock-induced \citep[][]{koy00,koy02,van07}, and 
magnetically-regulated formation of clouds \citep{fie00a,fie00b}, predict that 
these regions have a very filamentary structure at their early stages of evolution. Consistent with this idea, the filamentary IRDC G035.39-00.33
shows a relatively high number of quiescent cores 
(without H$_2$ shocked gas or 24$\,$$\mu$m sources), which 
are believed to be at a pre-stellar/cluster core phase 
\citep{cham09}. As a consequence, one should not expect to find 
a significant impact of outflow interaction within 
the cores and on their surroundings \citep[see e.g.][]{mar92,beu02}. 

The high-sensitivity maps of the SiO emission toward G035.39-00.33, 
however, reveal for the first time that SiO is widespread along an 
IRDC. Large-scale SiO
emission (with sizes ranging from 4 to 20$\,$pc) has also been
reported across the molecular clouds in the Galactic Centre 
\citep[][]{mar97,amo09}. In this case, the origin of these 
lines is different from that in
G035.39-00.33, because the SiO gas is highly turbulent
(linewidths of $\sim$60-90$\,$km$\,$s$^{-1}$). The large SiO fractional abundances
($\sim$10$^{-9}$) derived toward these regions are likely 
generated in fast shocks of supernova explosions, 
HII regions and Wolf-Rayet stellar winds \citep[][]{mar97},
and/or associated with X-ray or cosmic ray induced chemistry
\citep{amo09}. 
Besides the GC, the large-scale SiO emission observed toward 
this IRDC constitutes the largest SiO feature detected so far in a star 
forming region.
 
In Section$\,$\ref{res}, we have shown that the SiO line profiles 
measured toward G035.39-00.33 have two different components 
with different spatial distributions, kinematics and excitation. 
The first one consists of bright and compact SiO condensations 
(N, E and S) with broad line profiles in both SiO $J$=2$\rightarrow$1
and $J$=3$\rightarrow$2 transitions. From our 
excitation and fractional abundance analysis of the SiO lines, 
T$_{ex}$ and $\chi$(SiO) tend to progressively 
increase for larger velocities within the shock 
(from 6 to 40$\,$K, and from 
$\sim$10$^{-10}$ to $\geq$10$^{-8}$, respectively), as expected 
for shocked gas in molecular outflows \citep[][]{jim05,nis07}. 
Toward N and S, this idea is supported by the detection 
of 8$\,$$\mu$m sources, 24$\,$$\mu$m sources 
\citep{car09} and extended H$_2$ shocked gas \citep{cham09} 
closely 
associated with these condensations. In addition, the observed spatial extent of the 
broad SiO emission is only $\leq$30$''$ (i.e. $\leq$0.4$\,$pc at 2.9$\,$kpc),
which is consistent with those measured in outflows 
associated with high-mass protostars toward other IRDCs 
\citep{fall09}. We also note that this size is similar to the beam
size of the SiO $J$=2$\rightarrow$1 observations, for which the 
fractional abundances of SiO have been reproduced by our C-shock and 
sputtering models  (Section$\,$\ref{model}). 

For condensation E, we have seen that the broad SiO emission is not 
associated with any massive core in the region, and is clearly 
off the high extinction region \citep{rath06,but09}. 
In addition, no local maxima is seen in extinction or in other 
high-density molecular tracers in this region 
\citep[][Caselli et al. 2010, in preparation]{but09}. However, 
the IRAC 3-8$\,$$\mu$m and the MIPS 24$\,$$\mu$m fluxes of the source 
located 15$''$ north of the SiO peak are consistent with a 
15$\,$M$_\odot$ star in the ZAMS. Although one would not expect
strong outflow activity from such an evolved star, this source could be the 
origin of the broad SiO emission found in this condensation 
(Section$\,$\ref{nms}).  
Indeed, the SiO fractional abundances derived toward E 
are the largest measured in G035.39-00.33, and are consistent
with the idea that the SiO gas has been processed by a relatively 
strong shock 
(with $v_s$$\sim$30$\,$km$\,$s$^{-1}$; Section$\,$\ref{model}),
similar to those found in outflows associated with high-mass star 
formation \citep[see e.g. IRAS 20126+4104;][]{su07}. Therefore, although with the current data
it is not possible to clearly establish the origin of broad SiO 
toward condensation E, it is likely that this emission is 
related to shocked outflowing gas. 

The second SiO component detected toward G035.39-00.33, consists of extended and very narrow SiO emission that links 
condensations N, E and S, and which also spreads very faintly
toward the north and northwest of Core MM7/H6. From the low SiO 
fractional abundances 
derived for this component ($\sim$10$^{-11}$), the narrow SiO emission 
could be produced by the interaction of a 
large-scale low-velocity shock with $v_s$$\sim$12$\,$km$\,$s$^{-1}$ 
(Section $\,$\ref{model}) generated in the collision of two flows 
\citep{hen99,heit06,van07,hen08,heit09}. In G035.39-00.33,
this collision could have been produced by the 
interaction between the main filament \citep[as seen in the 
mass surface density map of][]{but09}, and the lower 
density filaments traced by C$^{18}$O (see Figure$\,$\ref{f3}). 
Since the time-scales required for SiO to freeze-out onto dust 
grains are relatively short \citep[i.e. from 5$\times$10$^3$$\,$yr 
to 5$\times$10$^4$$\,$yr for volume densities from 10$^{5}$ to 
10$^{4}$$\,$cm$^{-3}$; see Section$\,$6.2 in][]{mar92}, 
the dust grain processing 
event associated with this interaction would be relatively recent.
The narrow feature would then constitute a signature of the filament-filament 
collision or of previous accretion events produced onto 
the main IRDC filament.
 
This scenario is supported not only by the extended morphology of narrow
SiO, but by the fact that the coherent CO filaments observed toward G035.39-00.33 
resemble the molecular structures predicted in these models \citep{van07,hen08,heit09}.
In addition, the narrow SiO lines, specially toward the
north and northeast of Core MM7/H6, do not show any clear association with 
8$\,$$\mu$m or 24$\,$$\mu$m sources (Section$\,$\ref{narrow}). 
However, we note that the relatively small (2-4$\,$km$\,$s$^{-1}$) velocity difference
between the CO filaments compared to the shock velocity of $\sim$12$\,$km$\,$s$^{-1}$
required to produce the low SiO fractional 
abundances of $\sim$10$^{-11}$, does place constraints
on this scenario. This would require either that much of the relative
velocity between the colliding molecular gas is in the plane of the
sky or that much of the gas has already been decelerated in the
interaction. Detailed comparison with the results of numerical
simulations are required to assess the likelihood of these
possibilities.

As proposed by \citet{lef98} for the NGC1333 low-mass star forming region, 
the narrow SiO emission in G035.39-00.33 could arise from decelerated shocked gas 
associated with large-scale outflows driven by the 8$\,$$\mu$m and 24$\,$$\mu$m 
sources seen in the IRDC. This gas would have been decelerated by its 
interaction with a dense and clumpy surrounding medium \citep{lef98}. It is also possible
that narrow SiO is produced by material recently processed in the magnetic precursor 
of MHD shocks, as proposed for the L1448-mm outflow \citep[][]{jim04}. This 
idea is similar to that suggested by \citet{beu07}, for which narrow SiO
would be linked to the youngest jet/outflow objects present in their sample
of IRDCs. As discussed by these authors, 
narrow SiO lines are unlikely to be produced by an 
effect of outflow inclination with respect to the line-of-sight,
because this would lead to the detection of fewer outflows with broad 
line emission
than outflows with narrow SiO \citep[see Section$\,$3.2 in][]{beu07}.
Although any of the previous mechanisms could explain 
the narrow SiO lines in the ridge between condensations N, E and S, it seems unlikely for offsets (-10,70) and (-37,37), where the narrow SiO lines 
do not show a clear 
association with 8$\,$$\mu$m or 24$\,$$\mu$m sources (Section$\,$\ref{narrow}). 

Alternatively, the extended and narrow SiO emission toward G035.39-00.33
could be produced by a widespread and lower-mass population of protostars, 
compared to those powering condensations N, E and S. Some of these distributed 
protostars may be visible as the 8$\,$$\mu$m sources in Figure$\,$\ref{f1}, although 
we do see SiO emission from regions apparently devoid of such sources. In this scenario, 
beam dilution would then prevent us from detecting 
the broad SiO line wings expected to arise from these objects. 
Interferometric observations are thus needed to discriminate between i) the 
large-scale shock scenario, remnant of the IRDC formation process; ii) 
decelerated or recently processed gas in the precursor of MHD shocks in large-scale
outflows driven by the 8$\,$$\mu$m and 24$\,$$\mu$m sources; and iii) 
an undetected and widespread lower mass protostar population, as an origin of the widespread 
SiO emission in G035.39-00.33. 

Extended narrow SiO emission could also be produced by the UV photo-evaporation of the 
mantles of dust grains in photon dominated regions (PDRs) such as the Orion bar 
\citep[][]{sch01}. However, this mechanism seems unlikely in 
G035.39-00.33, because of the relatively low luminosity of the
region (it is observed as an infrared dark rather than bright cloud), and because 
the UV radiation field required to produce SiO fractional abundances similar to those observed 
in this cloud (of $\sim$10$^{-11}$), should be at least few hundred times the Galactic 
UV field \citep{sch01}. There is no evidence for sources capable of producing such 
an intense FUV field in this region, and even propagation of the much lower-intensity 
Galactic FUV radiation field into this cloud, would be strongly impeded by its high
extinction. Nor are cosmic ray induced UV photons \citep[][]{gre89} 
expected to play a key role in the formation of narrow SiO since the same 
UV field is generated in nearby quiescent dark clouds, where no SiO is
detected.
   
In summary, we report the detection of widespread 
(size of $\geq$2$\,$pc) SiO emission toward a very filamentary IRDC. This emission
presents two different components with different kinematics, excitation and spatial distributions.
The compact morphology, large SiO fractional abundances, and broad SiO 
line profiles observed toward N, E and S, indicate that these 
condensations are shocked gas in outflows associated 
with high-mass star formation. 
The second SiO component is extended 
along the filament and shows very narrow line profiles, low SiO abundances,
and lower excitation than the gas detected toward N, E and S. 
Although interferometric 
images are needed to clearly establish the origin of this 
emission, the properties of narrow SiO are consistent with 
i) a large-scale shock, remnant of the IRDC formation 
processes; ii) decelerated or recently shocked material in the precursor of shocks in 
large-scale outflows powered by 8$\,$$\mu$m and 24$\,$$\mu$m sources; or iii) an 
undetected and widespread population of lower mass protostars.

\section*{Acknowledgments}

We acknowledge the IRAM staff, and in particular H. Wiesemeyer, for the help 
provided during the observations. We also thank 
Prof. J. Mart\'{\i}n-Pintado for helpful discussions on the different 
mechanisms that can produce widespread SiO in star forming regions, and 
an anonymous referee for his/her careful reading of the manuscript.
J.C.T. acknowledges support from NSF CAREER grant AST-0645412.
FF acknowledges support by Swiss
National Science Foundation grant (PP002 -- 110504).
This effort/activity is supported by the European Community Framework
Programme 7, Advanced Radio Astronomy in Europe, grant agreement no.: 227290.

\end{document}